\documentclass[vecphys]{svmult}

\usepackage{graphicx}        
\usepackage[bottom]{footmisc}

\begin{document}

\title*{A Grand Challenge for PNe}
\author{Adam Frank\inst{}\, Orsola De Marco\inst{}\, Eric Blackman\inst{}\, Bruce Balick\inst{}}
\institute{Department of Physics and Astronomy, University of Rochester,
\\
Department of Astrophysics, American Museum of Natural History
\\
Department of Astronomy, University of Washington,
Seattle Wa, USA, 
}
%
%
\maketitle

\begin{abstract}

{The study of PN has
been confronting a growing list of dilemmas which have yet to find coherent resolution.  These issues are
both observational and theoretical and can be stated as a series of "facts" which can not, as of yet, be accounted for
via a single framework. We review these facts and propose a skeleton framework for developing a 
new understanding post-AGB stars, PPN and PN.  Our framework represents an attempt to articulate a  
a global perspective on the late stages of stellar evolution that can embrace both the nature of the 
central engine and the outflows they produce.  Our framework focuses on interacting binary central stars which 
drive collimated outflows through MHD processes.  We propose that the field of AGB/PN studies
now faces a "Grand Challenge" in articulating the observational systematics of these objects in a way that can
address issues related to binarity and magnetic shaping.  A theoretical Grand Challenge is also faced in the form
of integrated studies which can explicate the highly non-linear processes associated with MHD outflows
driven by interacting binaries.  These issues include the generation of magnetic fields via dynamo processes, the creation
of accretion disks, the dynamics of Common Envelope ejection and the creation of magnetized jets.}
 
\end{abstract}

\section{Method}
Since the onset of high resolution observational platforms, such as the Hubble Space Telescope, the study  of 
planetary nebula (PN)
 has confronted a growing list of paradoxes or dilemmas which have yet to 
find coherent resolution.  The dominant paradigm of single star evolution (Iben 1995) coupled with hydrodynamic 
interacting wind scenarios (e.g., Icke, Balick, Frank 1990, Frank \& Mellema 1994) is only able to explain a 
subset of the total mature PN population and fails entirely to explain the properties of pPN (Sahai \& Trauger 1998). 
This {\bf is an exciting time} as the new mechanisms being explored speak directly to some of the most important 
unsolved problems in astrophysics such as the nature of dynamo generated magnetic fields, accretion disk formation  
processes, collimated stellar explosions, stellar evolution in the context of binary stars and the interaction of jets 
and outflows with surrounding environments.

Progress  requires systematic exploration of both observational and theoretical issues. 
In this contribution, we  outline and justify a new framework for PN studies. We first provide a list of 
observational results or "facts" which have not, 
as of yet, been fully accounted for within traditional models. Next we provide a list of theoretical results 
or "facts" which demand   developing  a new global framework for PN outside of traditional single-star, 
radiation driven models.  In the final section we link observation and 
theory and propose a series of postulates that can serve as a straw-man for developing a paradigm for 
understanding PN and their related phenomena in a global evolutionary context.

\noindent{\bf Observational Facts}
\begin{enumerate}

\item{ {\bf PPN Momentum Excess:} The linear momentum observed in the majority of PPN outflows is higher than can be 
accounted for by
radiation driving from the central star even when multiple scattering is included.  These momentum excesses are typically
of order $1000$ or higher (Bujarrabal et al. 2001).}

\item{ {\bf Nebular Bipolarity:} All PPN and all very young PNe appear bipolar or multi-polar (Sahai \& Trauger 1998) even though 
all PN halos are spherical or mildly elliptical (Corradi et al. 2003).}

\item{ {\bf PPN and PN collimated structures:} Highly collimated structures, possibly jets exist in many PPN and PN.
(Balick \& Frank 2002 and references therein}

\item{ {\bf Post-AGB Binarity:} Virtually all post-AGB stars with dust {\bf tori} (which may be disks) have a binary companion.  Typical
periods are short enough to infer some form of interaction $100<P<1500$ days (van Winkel 2003).}

\item{ {\bf CSPN Binarity:} 
{\bf Close companions have been detected in $>10\%$ of central stars of PN  and $>10\%$ reveal  wide, non interacting companions, one system is a triple (Bond 2000; Ciardullo et al. 1999).}

\item{ {\bf Magnetic Field Detections:} Magnetic field measurements, while difficult, have been made
in both nebular systems and their central engines. Fields have been observed in 4 central stars as well 
as in a variety of AGB, pAGB/PPN and PN 
(Jordan et al. 2005; Vlemmings et al. 2006; Sabin et al. 2007).}

\item{ {\bf H-deficient central stars:} $15-20\%$ of all central stars are H-deficient. Most of these exhibit O and C dual 
dust chemistry.}

\item{ {\bf The PN Luminosity Function:} 
There is a ubiquitous population of bright PNe that comprise the same fractional brightness 
(the bright edge of the PNLF). This is in glaring conflict with predictions from population syntheses and stellar
evolution models which predict older galaxies to have overall fainter PNe (Marigo et al. 2004). 
Theory therefore predicts the PNLF should not be a good distance indicator, and yet observational 
practice puzzlingly demonstrates that PNLF is an accurate distance predictor.}

\item{ {\bf Bipolar PNe Scale Height:} Bipolar PNe tend to have a lower scale height and tend to be Type I. 
This is interpreted as meaning that more massive starts produce bipolar PNe (see E. Villaver's contribution, as this well 
known observation is in doubt in the LMC).}

\item{ {\bf PN Abundances} There is a discrepancy between PN abundances from optical recombination vs. forbidden lines. 
This discrepancy is alleviated by allowing small H-deficient clumps in the PN, however a physical reason for the existence of such clumps 
has yet to be found.}
}

\end{enumerate}

\noindent{\bf Theoretical Facts}
\begin{enumerate}
\item{ {\bf MHD Shaping:} The shapes of many bipolar PN and PPN can not be accounted for with the classic hydrodynamic 
Generalized Stellar Wind model.  MHD models, even those invoking weak magnetic fields, can produce a 
wide range of PN shapes (Garcia-Segura et al. 1997, 1999).}

\item{  {\bf Magnetic Fields and Momentum Excess:} Magneto-centrifugal launch (MCL) processes invoke a rotating central 
source and magnetic field (Pudritz 2004). These models can recover the high outflow momentum and energy observed in 
PN systems.  Thus MCL models can both launch and collimate PPN/PN outflows (Blackman et al 2001a,b, Frank \& Blackman 2004).}

\item{  {\bf Magnetic Fields and Short Acceleration Timescale:} The short acceleration timescales implied for many PPN 
flows can be accounted for via MCL models in which differential rotation in the central source produces 
strong toroidal field gradients which drive a "magnetic explosion" (Matt, Frank \& Blackman 2006).  The 
fragmentation of the expanding shell produced in such an explosion may provide a reasonable account for some 
multi-polar outflows (Dennis et al 2007).}

\item{  {\bf Magnetic Fields and Angular Momentum:} MCL models require relatively strong magnetic fields. These fields 
will require the action of a dynamo (Blackman et al 2001). Theoretical models show that it is likely that
such fields cannot survive across the AGB unless a binary companion acts to provide the angular momentum 	
lost in each dynamo cycle  (Soker 2006, Nordhaus, Blackman \& Frank 2006).} 

\item{  {\bf Binaries and Disks:} Circumbinary disks in PPN systems, now observed in a variety of cases, can form in 
a variety of ways including common envelope evolution (Nordhaus \& Blackman 2006) and Bondi accretion of the 
AGB wind (Soker \& Rappaport 2000).}

\end{enumerate}

\section{Towards a New Paradigm}
Below we provide postulates that follow from consideration of both the observational and theoretical facts. 

\begin{enumerate}
\item{  {\bf Almost all PN form from binary interactions including common envelopes:} {\it Required 
observations:} Systematic photometric variability (see D. Shaw's contribution) and radial velocity studies 
(see O. De Marco's contribution) of central stars of PPN and PN.  Goal: determine the frequency and period 
distribution of binaries. Systematic spectro-photometric monitoring study. Goal: determine the properties of
all known binaries (companion masses, orbital parameters, secondary irradiation characteristics) as well 
as those of their PNe (see B. Hrijvnak's and D. Frew's contributions).} 

\item{  {\bf Accretion disks will determine structure of many PPN:} {\it Required 
observations:} Systematic search for accretion disks in PN and PPN (by small scale photometric variability, 
X-ray detections [H emission lines likely contaminated by PN]).}

\item{  {\bf Binary interactions (avoiding common envelope) = binary pAGBs with circumbinary 
excretion discs:} {\it Required theoretical program:} Determine how intermediate separation binaries result in a massive, 
circumbinary disk. How do disks evolve in time? {\it Required observations:} Use systematic modeling of IR observations 
of systems known to have circumbinary dusty disks to determine properties.}

\item{  {\bf Binary interactions provide initial conditions for MHD launching:} {\it Required 
theoretical program:} A multi-dimensional numerical study of common envelope evolution and disk formation via a 
variety of mechanisms including mass transfer and wind capture and tracking energy deposition within the 
common envelope.}

\item{  {\bf Dynamo generated stellar fields will dominate evolution of PPN/PN:} {\it Required 
theoretical program:} Related to the point above this postulate requires systematic study of dynamo processes 
in AGB stars in both single stars and binary stars.  The potential for single stars to maintain strong fields 
via recirculation must be ascertained. Include development of multi-dimensional numerical simulations appropriate 
to stellar interiors.}

\item{  {\bf Magnetocentrifual launching accounts for PPN/PN morphologies:} {\it Proposed 
theoretical program:} A systematic study of MCL models in the context of PPN/PN central source configurations 
must be carried forward. The generation of outflows from transient MCL driven magnetic explosions as well as 
more continuous outflow produced by accretion disks must be explored.} 
\end{enumerate}

\noindent {\bf Conclusion} 
Our proposal that the majority of PN are shaped by binary stars and MHD processes holds promise for uniting the disparate observational and theoretical results cite above.  We note that many individual facets our proposal has been discussed before by other workers.  Our goal is to draw these together in a new synthesis which provide a direction for future coordinated work. We make this proposal with the understanding that much of the heavy lifting to prove, or disprove, our proposal has yet to be done. However it is noteworthy that the investigation of this paradigm will shed light not only on PN but also on other phenomena (YSOs, AGN, micro-Quasars, GRBs) currently at the forefront of astrophysical research.

%



\end{document}